# Coupling molecular spin qubits with 2D magnets for coherent magnon manipulation

*Sourav Dey[†1], Gonzalo Rivero-Carracedo[†1], Andrei Shumilin[1], Carlos Gonzalez-Ballestero[2]\* and José J. Baldoví[1]\**

[1] Instituto de Ciencia Molecular (ICMol), Universitat de Valencia, c/Catedrático José Beltrán, 2. Paterna 46980, Spain

[2] Institute for Theoretical Physics, Vienna University of Technology (TU Wien), Wiedner Haupstraße 8-10, 1040 Vienna, Austria

Email: j.jaime.baldovi@uv.es, carlos.gonzalez-ballestero@tuwien.ac.at

[†] Both contributed equally to this manuscript

ABSTRACT: Magnonics is an emerging field widely considered as a paradigm shift in information technology that uses spin waves for data storage, processing and transmission. However, the coherent control of spin waves in 2D magnets still remains a challenge. Herein, we investigate the interplay between molecular spins and magnons in hybrid heterostructures formed by [CpTi(cot)] and VOPc spin qubits deposited on the surface of the air-stable 2D van der Waals ferromagnet CrSBr using first principles. Our results show that different molecular rotation configurations significantly impact on qubit relaxation time and alter the magnon spectra of the underlying 2D magnet, allowing the chemical coherent control of spin waves in this material. We predict the feasibility of an ultrafast magnon-qubit interface with minimized decoherence, where exchange coupling plays a crucial role. This work opens new avenues for hybrid quantum magnonics, enabling selective tailoring through a versatile chemical approach.

**Introduction**

Magnonics is a rapidly growing field of research that deals with the storage, processing and transmission of information based on the use of spin waves (SWs), i.e. collective magnetic excitations in magnetic materials, instead of electric charges. This novel technology provides an alternative to electronics, enabling the development of nanoscale magnonic devices with lower power consumption and higher frequencies.[1,2] The recently born field of quantum magnonics[3] aims at exploring the potential of SWs –and their quanta, magnons– as components in hybrid quantum technologies. This is due to SWs' unconventional properties such as strong nonlinearity, tunability, or the ability to couple to almost every other degree of freedom,[4–6] which can be easily tailored when compared to other quantum particles such as phonons or photons. Similar to the latter, the interest of magnons for quantum platforms relies on the ability to manipulate and control their coupling to local quantum nodes, that is, qubits.[7] Several works have proposed methods to achieve this with superconducting qubits or solid-state spins, but they all focus on bulky ferrimagnetic samples.[8–13] The high damping and low scalability of these systems has sparked an active search for new platforms to implement the basic unit of hybrid quantum magnonics: a qubit-magnon interface.[14]

The discovery of long-range magnetic ordering in two-dimensional (2D) materials provides an unprecedented opportunity to study SWs on this class of systems. Among the family of layered van der Waals (vdW) magnetic materials, those that retain ferromagnetic order down to the monolayer limit such as $CrI_3$,[15] $Cr_2Ge_2Te_6$,[16] CrSBr,[17] $Fe_3GeTe_2$,[18] or $Fe_3GaTe_2$,[19] are at the forefront in research. These 2D magnets exhibit numerous advantages given that they represent the limit of miniaturization, have high flexibility, can be tuned through vdW stacking, and are compatible with silicon technology.[20,21] In this regard, a promising and unexplored route is the creation of hybrid molecular/2D heterostructures formed by spin qubits and 2D magnetic materials for coherent magnon control. Recently, some works have focused on the manipulation of SWs in bulk counterparts, especially nitrogen-vacancy (NV) centers in diamond, as they can be interfaced with other widely used excitations (microwave photons or phonons) with high quantum cooperativities and –crucially– they can be optically initialized at room temperature[12,22,23] and the impact of sublimable organic molecules and gas species on magnon transport in CrSBr has been evaluated.[24,25] However, the manipulation of SWs by coupling them to a spin qubit on 2D magnetic materials has not been reported to the best of our knowledge.

Among qubits, molecular spin qubits are particularly interesting. These are two-level systems based on coordination complexes of paramagnetic metal ions, which have been broadly investigated, and deposited on different substrates, showing strong hybridization with them.[26–28] In contrary to electron-spin based qubits, such as impurities in silicon[29] or nitrogen vacancies in diamond,[30] molecular spin qubits can couple with the substrate via magnetic exchange, which is much stronger than dipole-dipole interactions, allowing a more pronounced coupling to magnons. In addition, they exhibit higher tunability, i.e. different coordination geometries and ligands, thus opening new possibilities for the versatile coherent control of SWs.

In this work, we investigate the magnon emission –induced by a molecular spin qubit relaxation– in two hybrid molecular/2D magnetic heterostructures, formed by (i) [CpTi(cot)] (titanocene bis(cyclooctatetraenyl)) and (ii) VOPc (vanadyl phtalocyanine), deposited on the surface of the 2D ferromagnet CrSBr. Both molecules act as $S = 1/2$ qubits and have already been tested on different surfaces, preserving their spin state.[31–34] This has allowed them to efficiently tune superconductivity and give rise to Yu-Shiba-Rusinov states in superconducting lead.[35] On the other hand, CrSBr can be exfoliated down to the monolayer limit, exhibiting air-stability, high Curie temperature ($T_C$~146 K) and high energy magnons.[36–44] Through first principles calculations, we analyze the effects of these molecular spin qubits on the structural, electronic and magnetic properties of CrSBr monolayer, and particularly in terms of their coupling with magnons. This work fosters the exploration of novel frontiers at the spinterface regarding the coherent manipulation of SWs in 2D magnetic materials.

**Results and Discussion**

CrSBr is a layered material that crystallizes in an orthorhombic structure and *Pmmm* space group. The Cr atoms reside in a distorted octahedral coordination environment, $CrS_4Br_2$, within each layer, which leads to the energy splitting of the 3d orbitals into two sets, namely $t_{2g}$ ($d_{xy}$, $d_{xz}$ and $d_{yz}$) and $e_g$ ($d_{x^2-y^2}$ and $d_{z^2}$). The Cr atoms are linked via S and Br atoms along *a* axis and via S along the *b* and diagonal *ab* directions (see **Figure 1a, b**). According to our calculations, the lattice parameters of CrSBr monolayer are a = 3.545 Å and b = 4.735 Å, which are in close agreement with previous reports.[45,46] As a molecular counterpart, we use (i) [CpTi(cot)], which is an organometallic complex in which the metal atom is sandwiched between an eight-membered ring, $\eta^8$-cyclooctatetraene ($cot^{2-}$), and a five-membered ring, $\eta^5$-cyclopentadienyl ($Cp^-$),[47] and (ii) VOPc, which is a nonplanar metal-phthalocyanine complex where the vanadyl ion ($VO^{2+}$) is coordinated to the four nitrogen atoms of the phthalocyanine ($Pc^{2-}$) ring (see **Figure 1c, d**).[48]

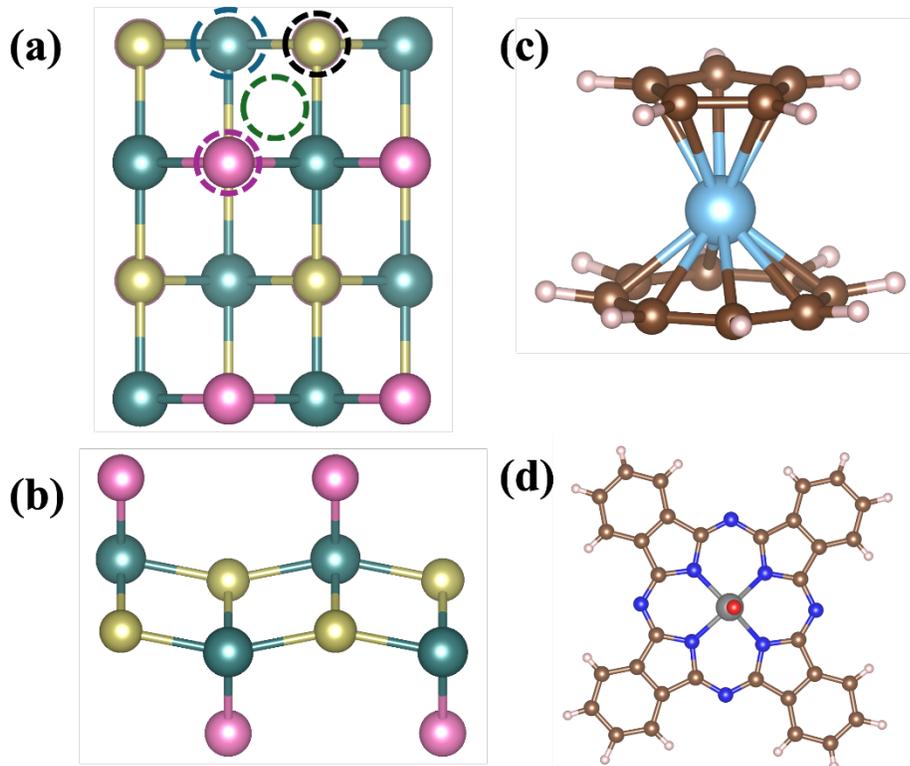

**Figure 1**: (a) Top view and (b) side view of CrSBr monolayer. The four adsorption sites such as top of Cr, Br, S and hollow are indicated with blue, purple, orange and green circles, respectively. Molecular structure of (c) [CpTi(cot)] and (d) VOPc. Colour code: Cr (green), Br (pink), S (yellow), Ti (cyan), V (silver grey), O (red), N (blue), C (brown) and H (white).

First, we study [CpTi(cot)] and VOPc molecules in gas phase to determine their electronic structures. DFT calculations reveal the splitting of the 3d orbitals into non-bonding, bonding and anti-bonding sets, in both molecules. The non-bonding $d_{z2}$ orbital in [CpTi(cot)] and $d_{xy}$ orbital in VOPc are the singly occupied molecular orbitals (SOMO), which are the magnetic orbitals in each molecule (see **Figures S1** and **S2** in the **Supporting Information** for details).

To investigate the effect of qubit adsorption on the magnetic properties of CrSBr, we design [CpTi(cot)]@CrSBr and VOPc@CrSBr heterostructures with different possible adsorption geometries, using 4 × 4 and 6 × 6 CrSBr supercells, respectively. In order to consider the molecular configuration on the substrate, we study three different orientations for [CpTi(cot)]: *standing$_{cot}$*, *standing$_{Cp}$* and *lying* (**Figures 2** (top) and **S3** (top)). On the other hand, we study two possible

VOPc orientations on CrSBr: *oxygen-up* and *oxygen-down* (**Figures 2** (bottom) and **S3** (bottom)). Four different adsorption sites on CrSBr monolayer for both systems are considered, placing the metal atom of each qubit sitting on top of Br, Cr, S and hollow, as shown in **Figure 1a**.

The calculated equilibrium distance between [CpTi(cot)] and CrSBr is 2.75-3.09 Å depending on the orientation (see **Figure S4** and **Table S1**). Note that, among the three orientations of [CpTi(cot)], *standing$_{cot}$* is the closest case (2.75 Å), exhibiting stronger interaction with the substrate compared to *standing$_{Cp}$* and *lying* orientations. In the case of VOPc, the largest interaction is observed for the *oxygen-up* orientation, given that the aromatic ring of the Pc is parallel to the surface. This allows a larger hybridization when compared to *oxygen-down* (**Table S2**). The molecule-substrate distance is around 3.0 Å, which also indicates a physisorption process. One can observe that a larger deformation occurs in bonding parameters in [CpTi(cot)] when compared to those of VOPc after adsorption. This implies that the former interacts more strongly with the substrate (**Tables S1** and **S2**). Subsequently, we calculate the adsorption energy ($E_{ads}$) to study and compare the stability of the different heterostructures. Our calculations indicate that the *standing$_{cot}$* orientation ($E_{ads}$ = -0.94 eV, **Table** S3) is the most stable heterostructure for [CpTi(cot)], followed by *standing$_{Cp}$* ($E_{ads}$ = -0.67 eV, **Table** S3) and *lying* ($E_{ads}$ = -0.53 eV, **Table** S3) orientations, respectively. All these magnitudes reveal a physisorption process governed by vdW interactions. In general, the difference in the adsorption energy values for both different adsorption sites and orientations is small, which limits the prediction of a preferential adsorption configuration (**Table S3**) for the molecules. This is consistent with the observed behavior for [CpTi(cot)] on Au(111) surface.[31] For VOPc, the computed adsorption energy is 0.7 eV larger with respect to that of [CpTi(cot)] (**Table** S4). In particular, the *oxygen-up* orientation on top of the Cr site is found to be the most stable heterostructure ($E_{ads}$ = -1.65 eV, **Table** S4). In contrast, for *oxygen-down* orientation, the adsorption energy was estimated to be 1.1 eV smaller compared to *oxygen-up* orientation (**Table S4**). Like [CpTi(cot)], the adsorption energy differs by less than 0.1 eV for the four adsorption sites (**Table S4**).

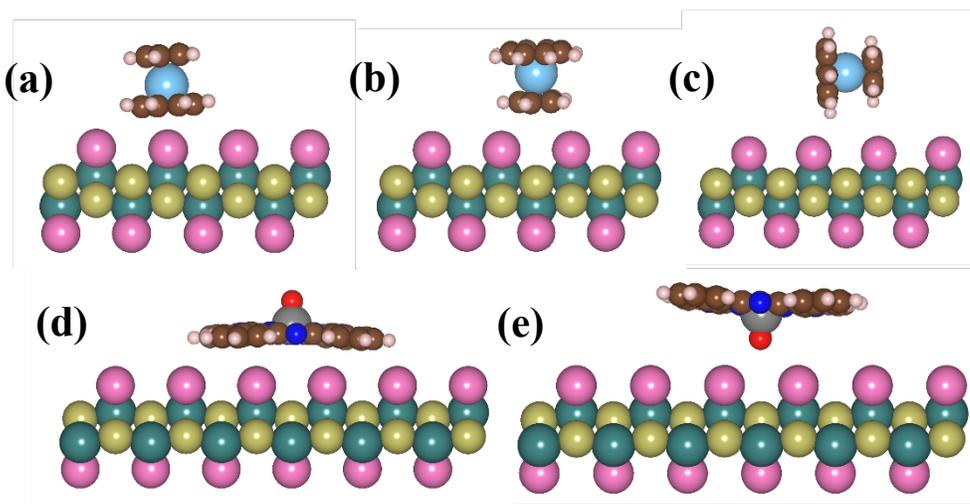

**Figure 2**: (Top) Side views of the optimized adsorption geometries of [CpTi(cot)] on the most stable site of CrSBr in three orientations: (a) *standing$_{cot}$*, (b) *standing$_{Cp}$*, and (c) *lying*. (Bottom) Side views of the optimized adsorption geometries of VOPc on the most stable site of CrSBr, showing (d) *oxygen-up* and (e) *oxygen-down* configurations.

Then, we perform a Bader charge transfer analysis[49] to study the electron density flow at the interface for the most stable adsorption site in both hybrid heterostructures. Our calculations evidence the transfer of electrons from the molecule to the substrate for all molecular orientations and adsorption sites. For [CpTi(cot)], the largest charge transfer (0.46e) is estimated for *standing$_{cot}$* orientation (which presents the largest adsorption energy), followed by 0.37e for both *standing$_{Cp}$* and lying orientations. The charge density difference (CDD) plot suggests charge depletion in the Cp and cot rings, whereas the charge accumulation occurs in the Br atoms of the substrate. This is due to the high electronegativity of the latter (see **Figures 3a,b** and **S5**). In the case of VOPc, a charge transfer of 0.24e and 0.16e is calculated for *oxygen-up* and *oxygen-down* orientations, respectively. Here, charge depletion occurs from the phthalocyanine macrocycle (**Figures 3d,e** and **S6**). Then, to understand the effect of electron delocalization on the magnetic properties, we calculate the spin density of the heterostructure and compare it with the isolated molecule (**Figures 3c,f** and **S7**). Our calculations on [CpTi(cot)]@CrSBr reveal a 46% reduction in the magnetization of Ti, while the average magnetization of Cr in the substrate increases by 0.6% due to electron delocalization from the molecule to the substrate (**Figures 3e**, **S8** and **S9**). For VOPc, the magnetization of V remains almost unperturbed in both *oxygen-up* and *oxygen-down* orientations, and consequently, shows negligible spin delocalization with the substrate (**Figure 3f** and **Figure S6**, **S10**).

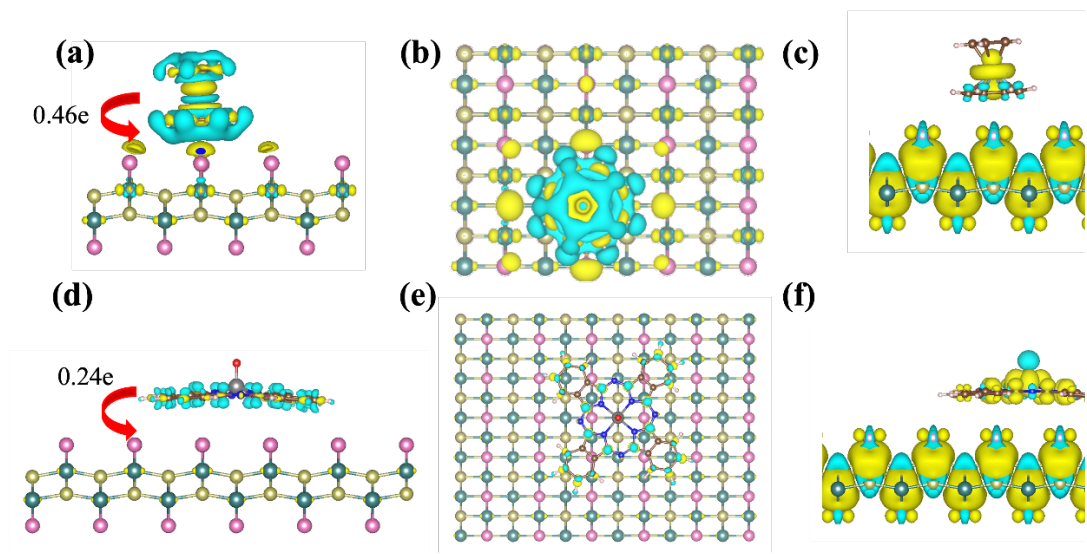

**Figure** 3: Charge density difference (CDD) plots of the CrSBr heterostructure for the *standing$_{cot}$* orientation of [CpTi(cot)]: (a) side view and (b) top view, and for the *oxygen-up* orientation of VOPc: (d) side view and (e) top view. Yellow and blue regions indicate charge accumulation and depletion, respectively, with an isosurface value of 0.0005 eVÅ$^{-3}$. Spin density distribution in the CrSBr heterostructure for (c) the *standing$_{cot}$* orientation of [CpTi(cot)] and (f) the *oxygen-up* orientation of VOPc, with an isovalue of 0.001 eVÅ$^{-3}$.

Furthermore, we analyze the effect of molecular adsorption on the projected density of states. Our results show the appearance of new localized states, while the intrinsic band gap of the pristine CrSBr remains unchanged (**Figure S11** and **S12**). For both molecules, the discrete levels are not broadened, suggesting a physisorption process where the interfacial interactions are dominated by vdW forces. The calculated band structure and density of states show a stronger mixing of the molecular orbitals of [CpTi(cot)] with the substrate when compared to VOPc, which is in accordance with its more pronounced charge transfer and spin density. Among the three orientations of [CpTi(cot)], *standing$_{cot}$* shows stronger hybridization with the substrate (**Figure S11b**). We also calculate the projected density of states for the molecule by removing the substrate atoms and comparing it with the hybrid heterostructure. For [CpTi(cot)], we find that the non-bonding $d_{z^2}$ orbital of Ti has been shifted from ca. -2.0 eV to the CBM in the hybrid heterostructure (**Figures S13-S15**). On the other hand, we observe that the 2p orbitals of carbon are shifted to the Fermi level after adsorption (**Figures S16** and **S17**). This can be attributed to the charge transfer from the 3d orbital of Ti in [CpTi(cot)] and the 2p orbitals of carbon in VOPc to the CrSBr monolayer.

Then, we investigate the effect of the adsorption of the qubit molecule on CrSBr magnetic exchange interactions. Magnetic couplings in pristine CrSBr are usually modelled by the three nearest-neighbor exchange interactions, namely $J_1$, $J_2$ and $J_3$ (**Figure 4a**). $J_1$ represents the magnetic exchange between Cr atoms along the *a* axis, $J_2$ denotes the exchange interaction along the *ab* diagonal direction and $J_3$ accounts for the exchange interaction along the *b* crystallographic direction (see characteristic angles in **Table S5** and **Figure S18** in **Supporting Information**). The adsorption of the molecular spin qubits on the 2D material leads to an additional exchange ($J_4$), which takes place between the spin of the molecule and nearest-neighbors in the substrate (**Figure**

**4b**). These exchange interactions are determined by mapping the energies of different magnetic configurations (see **Figures S19** and **S20**) into an isotropic Heisenberg Hamiltonian of the form:

$$\hat{H} = -\sum_{ij} J_{ij} \hat{s}_i \hat{s}_j \quad (1)$$

where $J_{ij}$ represents the exchange interaction between two different spins ($\hat{s}_i$ and $\hat{s}_j$). Additionally, the qubit energy is calculated as: $E_Q = S_{Cr} S_Q N_N J_4$, where $S_{Cr}$ represents the spin of Cr, which is 3/2; $S_Q$ the spin of the qubit, which is 1/2 for both molecules; and $N_N$ the number of nearest-neighbours Cr atoms to the spin qubit. The dipole-dipole interaction of the spin qubit with CrSBr ($E_{dd}$) was calculated from optimized atomic coordinates. The estimated magnetic exchange interactions for $J_1$-$J_4$, qubit energies and dipole-dipole interactions are reported in **Table 1**.

**Table 1**. Isotropic magnetic exchange parameters (meV), qubit energies ($E_Q$ (meV)) and dipole-dipole energies ($E_{dd}$ (meV)) for different molecular orientations of [CpTi(cot)] and VOPc on CrSBr. $J_1$-$J_3$ for pristine CrSBr are also reported for comparison.

|  | | [CpTi(cot)] | | | VOPc | |
|---|---|---|---|---|---|---|
|  | CrSBr | *standing<sub>cot</sub>* | *standing<sub>Cp</sub>* | *lying* | *oxygen-up* | *oxygen-down* |
| $J_1$, meV | 3.294 | 3.173 | 3.117 | 3.122 | 2.993 | 2.976 |
| $J_2$, meV | 3.994 | 4.038 | 4.044 | 4.044 | 4.003 | 4.001 |
| $J_3$, meV | 2.328 | 3.123 | 3.014 | 3.064 | 2.303 | 2.256 |
| $J_4$, meV |  | 0.225 | 0.098 | 0.010 | -0.009 | 0.002 |
| $E_Q$, meV |  | 1.802 | 0.782 | 0.080 | 0.144 | 0.020 |
| $E_{dd}$, meV |  | 0.018 | 0.013 | 0.012 | 0.018 | 0.009 |

The calculated magnetic exchanges for pristine CrSBr are 2.9 meV ($J_1$), 4.0 meV ($J_2$) and 2.2 ($J_3$) meV and agree very well with experimental values.[40] These exchange interactions are strongly modified in the [CpTi(cot)]@CrSBr hybrid heterostructure compared to the VOPc one. The strongest effect is a 34% increase of $J_3$ by the adsorption of [CpTi(cot)] in the *standing*cot orientation. This change in magnetic exchange after adsorption stems from (i) substrate distortion, (ii) charge transfer from molecule to substrate, and (iii) molecule-substrate exchange interaction (see **Table S7** and **Supporting Information** for details). On the other hand, $J_4$ depends on both the orientation of the molecule and adsorption site on CrSBr, as we show in **Table 1**. According to our simulations, $J_4$ is 0.225 meV for *standing$_{cot}$* orientation of [CpTi(cot)] and -0.009 meV for *oxygen-up* orientation of VOPc. This molecule-substrate exchange interaction can occur through (i) direct exchange between the metal center and the 3d orbitals of Cr or (ii) indirect coupling via 2p orbitals of coordinated atoms. In [CpTi(cot)], indirect exchange leads to antiferromagnetic coupling, as seen in the spin density (**Figure 3c**), making direct interaction responsible for ferromagnetic coupling with CrSBr. For VOPc, antiferromagnetic coupling arises from indirect interaction via nitrogen 2p electrons, which is consistent with the calculated spin density (**Figure 3f**). Notably, in all cases $E_Q$ is much larger than $E_{dd}$, showing a good interplay between the molecular spin qubits and the magnetic properties of CrSBr.

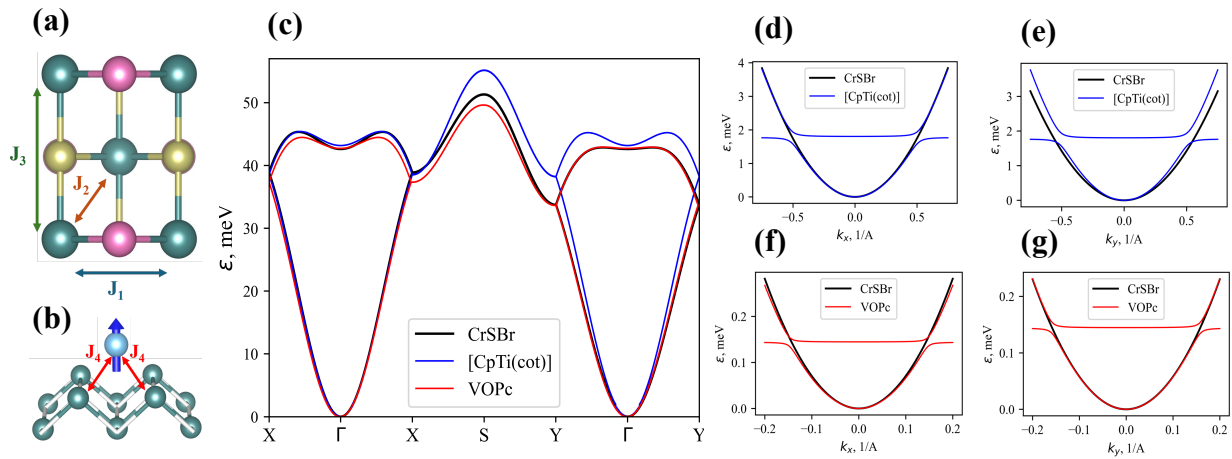

**Figure** 4: Schematic representation of the nearest-neighbor exchange interactions (a) $J_1$, $J_2$ and $J_3$ for CrSBr, and (b) $J_4$ between molecular spin and the nearest Cr atoms from the substrate. (c) Magnon spectrum of pristine CrSBr and the [CpTi(cot)]@CrSBr and VOPc@CrSBr heterostructures. (d-g) Insets of low-energy magnons near Γ-point.

The magnon spectra for [CpTi(cot)] (*standing$_{cot}$*) and for VOPc (*oxygen-up*) are shown in **Figure 4**. **Figure 4c** presents their unfolded acoustic and optical magnon bands, as well as the magnon spectrum of pristine CrSBr. The most notable effect is the increase in optical magnon energy induced by the [CpTi(cot)] qubit. The changes in the magnon spectra shown in **Figure 4c** primarily stem from modifications of $J_1$, $J_2$ and $J_3$, rather than direct exchange interactions with the qubits. However, the molecular spins introduce an additional magnon band, absent in pristine CrSBr. To examine this feature, we plot the magnon spectra near the Γ-point without unfolding, as shown in **Figures 4d–g**. This additional band remains relatively flat due to the lack of direct qubit-qubit exchange, as qubits interact only via magnons in CrSBr. However, anti-crossing occurs between this "qubit band" and the acoustic magnons of the 2D magnet, leading to strong hybridization of magnons and spin qubits at specific points in the momentum space. The weak indirect qubit-qubit interaction suggests that, in realistic scenarios with randomly distributed qubits, their spins may become localized. Nevertheless, qubit-magnon hybridization persists, manifesting as enhanced magnon scattering and magnon absorption by the qubits.

While a qubit layer modifies the magnon spectrum, individual spin qubits act as magnon emitters (**Figure 5a**). The qubit-magnon interaction is significantly stronger than qubit-phonon coupling, which typically results in qubit relaxation times on the order of 1μs or larger.[50–52] Consequently, magnon emission dominates the relaxation process. Using our spin Hamiltonian framework, we analyze qubit-magnon interactions and find that qubit relaxation is characterized by relaxation times of 15.7ps, 82.3ps and 7.19ns for the *standing$_{cot}$*, *standing$_{Cp}$* and *lying* orientations of [CpTi(cot)], respectively, and 2.46ns for VOPc (*oxygen-up*). Relaxation curves for various qubits and orientations are shown in **Figure 5b**. In **Figures 5c-j**, one can observe that qubits relax emitting a single magnon pulse with full spatial coherence, represented by magnon occupation probabilities. We perform these simulations for *standing$_{cot}$* and *standing$_{Cp}$* orientations at different post-relaxation times, as well as for the less stable cases, i.e. *lying* [CpTi(cot)] and *oxygen-down* VOPc (**Figure S22**). Initially, while the qubit remains significantly excited, the magnon state appears as a localized cloud around the qubit. Over time, a ring-like structure emerges, expanding with the magnon group velocity. In the *lying* orientation of the [CpTi(cot)] qubit and for VOPc, the lower energies lead to magnon evolution on the nanosecond timescale and micrometer length scale, as detailed in the **Supporting Information**. The shape of the coherent magnon state depends on the qubit relaxation time and the group velocity of magnons matching the qubit energy. Faster qubit relaxation results in a smaller, more rapidly forming ring with a thinner structure.

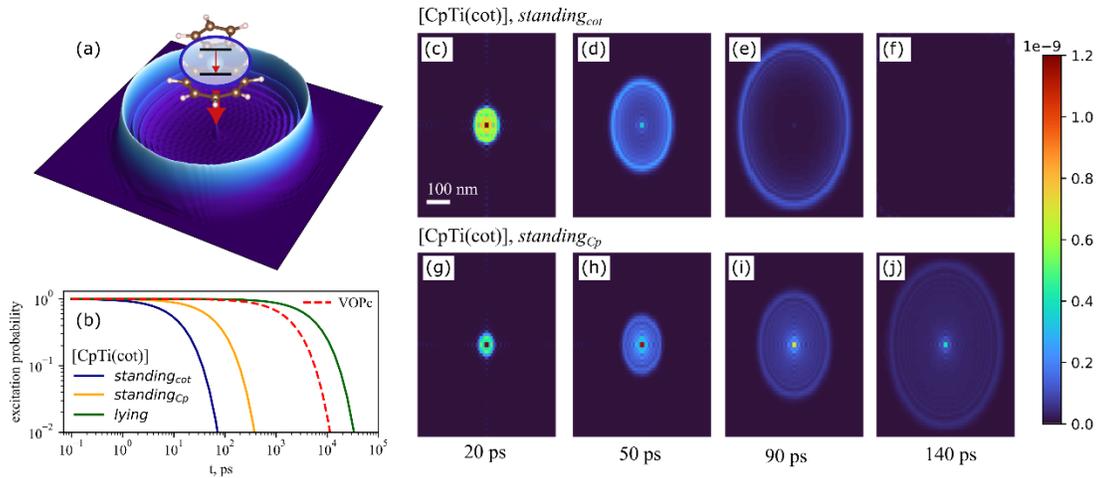

**Figure** 5: Molecular spin qubit relaxation. (a) Representation of the spin qubit relaxing by emitting a single-magnon pulse, (b) the time-dependent excitation probability of different qubits. (c-f) calculated shapes of the single magnon pulse emitted by [CpTi(cot)] qubit in *standing$_{cot}$* orientation at different post-relaxation times. (g-j) calculated shapes for the relaxation of [CpTi(cot)] in *standing$_{Cp}$* orientation.

The fast magnon dynamics is a core asset for any quantum protocol involving qubits and propagating fields, enabling high-speed quantum state preparation and processing. Furthermore, this is a core requisite for magnons to be useful in hybrid quantum platforms, as their high losses (compared to e.g. photons) demand ultrafast protocols to allow using the unconventional magnon properties at minimized quantum coherence loss. The fast dynamics in the studied hybrid heterostructures is achieved by magnetic exchange coupling, which can be tuned by charge transfer from molecule to substrate. Therefore, an optimal way of increasing the molecule-substrate exchange coupling may be the introduction of an alkyl group in any of the 8-member or 5-member ring of [CpTi(cot)], as this will enhance the donor character of these rings. In VOPc, weak coupling arises from an unpaired electron in the $d_{xy}$ orbital, which lacks strong overlap with the substrate and bonded atoms. This results from the strong crystal field of the V=O bond, which lifts orbital degeneracy. Selecting qubits with weaker axial ligand fields could enhance molecule-substrate coupling. Furthermore, to strengthen molecule-substrate coupling, qubits with S = 1/2 organic radicals like TEMPO (2,2,6,6-Tetramethylpiperidine-1-oxyl) and nitronylnitroxide are promising candidates, as they can be engineered for chemisorption onto the substrate through appropriate chemical modifications.[26,53–55] Additionally, Br defects in CrSBr could facilitate electron donation from the molecule due to the electron deficiency in the substrate.

In summary, we investigate the electronic structure, magnetic properties and quantum magnon dynamics of [CpTi(cot)] and VOPc spin qubits deposited on single-layer CrSBr. Besides tuning the magnon band structure of the 2D magnet, our calculations predict that qubits relax emitting an ultrafast single magnon pulse with full spatial coherence. We observe the most pronounced effects in the hybrid heterostructure formed by [CpTi(cot)] on CrSBr, in which we demonstrate that the relaxation time can be selectively tuned from 16 ps to 7 ns in as a function of molecular orientation. Additionally, we provide a detailed microscopic understanding of the qubit-substrate coupling, shedding light on the rational exploitation of coherent magnon dynamics in hybrid molecular/2D magnetic heterostructures. This work paves the way to hybrid quantum platforms that harness magnons in an ultrafast regime, thus leveraging their unconventional properties while minimizing decoherence caused by magnon loss.

**Computational Details**

The geometry optimization of the isolated molecule was performed with density-functional theory (DFT) calculations with the ORCA (v. 5.0.3) package.[56] The DFT calculations were performed with PBE0 functional and def2- TZVP basis set.[57] The frontier orbitals were calculated with the def2/J auxiliary basis set and RIJCOSX approximation.[58]

All the spin-polarized DFT calculations on the CrSBr monolayer were carried out using the QUANTUM ESPRESSO package.[59] The Perdew–Burke–Ernzerhof (PBE) exchange-correlation functional was employed in the framework of generalised gradient approximation (GGA).[60] The lattice parameters and atomic coordinates were fully relaxed using Broyden–Fletcher–Goldfarb–Shanno (BFGS)[61] algorithm within the convergence criteria of $1 \times 10^{-3}$ Ry/au forces for each atom and $1 \times 10^{-4}$ Ry energy difference between two consecutive relaxation steps. We have used 60 and 600 Ry for kinetic energy and charge density cutoffs for the expansion of the electronic wavefunction. All the pseudopotentials were taken from the solid-state ultrasoft library of the Quantum ESPRESSO package. To avoid unphysical interlayer interactions along $c$ directions, an 18 Å vacuum spacing was employed. A G-centered $8 \times 8 \times 1$ k-point Monkhorst–Pack grid was used to integrate the Brillouin zone.

Thereafter, we have constructed $4 \times 4$ and $6 \times 6$ supercells for the [CpTi(cot)]@CrSBr and VOPc@CrSBr heterostructures to ensure a minimum distance of 8 Å between the neighbouring molecules as we are interested in investigating the individual effect of qubits on the magnetic properties of CrSBr. We have employed Grimme-D2 dispersion corrections in our calculations to take into account the vdW interactions between qubit and the substrate.[62] To take into account the electron correlations in 3d orbitals of Cr and V, we have used an $U_{eff}$ =3eV (U=4eV and $J_H$ =1eV) using Liechtenstein's formulation for both of them.[63]

We have only relaxed the atomic coordinates for the heterostructure as the effect of single qubit on the lattice parameter of the substrate will be minimal. Here, the Brillouin zone was sampled by Γ-centred 4×4×1 (2×2×1) k-point grid for [CpTi(cot)]@CrSBr (VOPc@CrSBr). We have calculated the adsorption energy as follows,

$$E_{ads} = E_{CrSBr+qubit} - E_{CrSBr} - E_{qubit} \qquad (1)$$

where $E_{CrSBr+qubit}$, $E_{CrSBr}$, $E_{qubit}$ correspond to the energies of the relaxed geometry of hybrid heterostructure, pristine CrSBr, and isolated qubit, respectively.

The charge transfer between the qubit and CrSBr was calculated using Bader charge population analysis.[49] To visualize the charge density difference (CDD), the charge density of isolated qubits and pristine CrSBr monolayers were subtracted from the total charge density of the hybrid heterostructure using the following formula,

$$\Delta\rho = \rho_{CrSBr+qubit} - \rho_{CrSBr} - \rho_{qubit} \qquad (2)$$

where, $\rho_{CrSBr+qubit}$ and $\rho_{CrSBr}$ corresponds to the total electron densities of monolayer CrSBr with and without qubit respectively and $\rho_{qubit}$ denotes the electron density of the isolated qubit. It is worth mentioning that separate qubit and CrSBr monolayer should have the same distorted geometry as in the hybrid heterostructure.

The magnetic exchange interactions ($J_1$, $J_2$ and $J_3$) in the CrSBr monolayer, along with CrSBr — qubit exchange interaction $J_4$, are determined by mapping the total energies of five distinct magnetic configurations, one ferromagnetic and four antiferromagnetic, onto a classical Heisenberg Hamiltonian.

Magnon spectra are computed using the linear approximation of the Holstein–Primakoff boson expansion,[64] with calculations performed using the RAD-tools software.[65] The unfolding procedure is carried out by analyzing magnon wavefunctions, as detailed in **Supporting Information**. Magnon relaxation rates and probability distributions are obtained within the Weisskopf-Wigner framework[48], justified by the efficient magnon transport in CrSBr, which ensures Markovian qubit relaxation. The corresponding quantum-mechanical calculations are provided in **Supporting Information**.

**Author contributions**

[†]S.D. and [†]G.R.C. contributed equally. The manuscript was written through contributions of all authors. All authors have given approval to the final version of the manuscript.

**Notes**

The authors declare no competing financial interest.

**Funding Sources**

The authors acknowledge the financial support from the European Union (ERC-2021-StG-101042680 2D-SMARTiES, FETOPEN SINFONIA 964396 and Marie Curie Fellowship SpinPhononHyb2D 10110771), the Spanish MICINN (Excellence Unit "María de Maeztu" CEX2019-000919-M) and the Generalitat Valenciana (grant CIDEXG/2023/1). G. R. C. thanks the University of Valencia (grant Atracció de Talent INV23-01-13). C.G.B. acknowledges the Austrian Science Fund FWF for the support with the project PAT-1177623 "Nanophotonics-inspired quantum magnonics". The calculations were performed on the HAWK cluster of of the 2D Smart Materials Lab hosted by Servei d'Informàtica of the University of Valencia